\documentstyle[epsfig,aps]{revtex}

\begin{document}

\draft

\def\lsim{\lower.5ex\hbox{$\; \buildrel < \over \sim \;$}}
\def\gsim{\lower.5ex\hbox{$\; \buildrel > \over \sim \;$}}


\title{Neutrino asymmetry around black holes: Neutrinos interact with gravity}
\author{Banibrata Mukhopadhyay\thanks{bmukhopa@cfa.harvard.edu}\\ 
{\sl Theory Division, Harvard-Smithsonian Center for Astrophysics,
60 Garden Street, MS-51, Cambridge, MA 02138, USA} 
}

\maketitle
\baselineskip = 18 true pt
\vskip0.3cm
\setcounter{page}{1}

\def\ch{\lower-0.55ex\hbox{--}\kern-0.55em{\lower0.15ex\hbox{$h$}}}
\def\lh{\lower-0.55ex\hbox{--}\kern-0.55em{\lower0.15ex\hbox{$\lambda$}}}       
\def\n{\nonumber}

\begin{abstract}

Propagation of a fermion in curved space-time generates a gravitational interaction due 
to coupling of its spin with the space-time curvature connection. This gravitational 
interaction, which is an axial-four-vector multiplied by a four gravitational vector 
potential, appears as a CPT violating term in the Lagrangian which generates an opposite 
sign and thus asymmetry between the left-handed and the right handed partners under 
the CPT transformation. In the case of neutrinos, this property can generate a neutrino 
asymmetry in the Universe. If the background metric is of the rotating black hole, 
i.e. the Kerr geometry, this interaction for the neutrino is non-zero. Therefore, the 
dispersion energy relations for the neutrino and its anti-neutrino are different 
which give rise to the difference in their number densities and the neutrino 
asymmetry in the Universe in addition to the known relic asymmetry.

\end{abstract}

\vskip1.0cm
\vskip2cm


\section{Introduction}

The generation of neutrino asymmetry, i.e., the excess of neutrinos over anti-neutrinos,
in early Universe is a well known
fact. This  essentially arises due to the lepton number asymmetry, e.g.
via the Affleck-Dine  mechanism \cite{admcdonald}, in the early Universe.
A large neutrino asymmetry in the early Universe
can have interesting effects on various cosmological phenomena like
big-bang nucleosynthesis and cosmic microwave background \cite{cosmoeffects}.
Massive neutrinos with large asymmetry can also offer to explain existence
of cosmic radiation \cite{fargion,weiler} with energy greater than GZK cutoff \cite{gzk}.
Apart from such asymmetry arising in the early Universe, one can always ask
whether there is a possibility of the neutrino asymmetry arising when the 
Universe has cooled down, let us say in the present era. In this paper
we present one such scenario when neutrinos are propagating around
Kerr black holes. 

Since long, propagation of test particles with some inherent structure in curved space-times
has been of keen interest at both the classical and quantum realms. 
A spinning test particle when propagates in the gravitational field, its
spin couples with the connection of the background space-time and produces an
interaction term \cite{pap51,dixon,ads}.
A similar coupling effect gets transferred to the phase factor at the
quantum mechanical level leading to an interesting geometrical phase shifts (see e.g. \cite{jeeva}).
This interaction between the spin of the particle and the spin connection of 
the background field is analogous to 
that of the electric current with the vector potential in the case 
of the electromagnetic field.
The way electro-magnetic connection serves as a gauge field, 
in a similar manner, in
case of fermions in curved space, the gravitational interaction gives rise to
some sort of gauge field \cite{m00}.

The propagation of fermions in the curved space-time is well studied in past
by several authors (e.g. \cite{m00,p71,chandra76,p80,lalak,bd,sw,mmp02}).
The interaction term does not seem to preserve CPT \cite{mpla}, and is similar to the 
effective CPT as well as Lorentz violating terms as described in other contexts in previous 
works (e.g. \cite{ck,d99,bklr}).
Therefore the interaction due to the curvature coupling of the spinor will give rise
 to opposite sign for a left-handed and right-handed field which for the
case of neutrinos can lead to an asymmetry. A preliminary result, based on this scenario, 
was already reported by us \cite{mpla,pram}. Also a similar asymmetry was noted by
Ahluwalia-Khalilova \cite{ahl1}. In connection with this fact, the gravitationally-induced 
neutrino oscillations were studied \cite{ahlbur,ahl2,kk,wud}. 
Later, the Lorentz and CPT violation scenario was
addressed by Kosteleck\'y in the context of Riemann-Cartan space-times \cite{kosgrav}.

In this paper, we elaborate our earlier results\cite{mpla}, describing various
aspects, in detail and showing its application around black holes.
We show that such a neutrino asymmetry  can arise even in the 
present epoch like in the black hole space-times. In fact, we would show, it
is just the form of the background metric which is responsible for such an effect. If 
the background metric satisfies a particular form which we discuss
below and if the temperature of the bath is large enough, then the favorable
conditions for neutrino asymmetry exist. 
In this connection, obviously the Dirac equation and corresponding Lagrangian in curved background comes
into the picture. Under curved space-times Dirac spinors can break the Lorentz invariance 
in the local frame which provide a 
background where the ordinary rules of quantum field theory, e.g. CPT invariance, can 
break down. It is seen that coupling between the fermionic spin and curvature
of the space-time gives rise to an extra interaction term in the Lagrangian apart from
free part, even if no further interaction is there. This interaction term
does not preserve CPT and Lorentz symmetry.

The basic requirement to generate the neutrino asymmetry by this mechanism 
is that the background metric 
should deviate from spherical symmetry, like that of a Kerr black hole.
If the black hole is chosen to be non-rotating (e.g. Schwarzschild type), 
then the CPT violating interaction term
disappears and the neutrino asymmetry is ruled out. In next section, we give the mathematical
formalism, which clearly shows the neutrino asymmetry is possible to generate in present era. 
In \S 3, we give an example where this asymmetry can arise in the black hole space-time.
At last, in \S 4, we make  conclusions.

\section{Formalism to produce neutrino asymmetry}

The general Dirac Lagrangian density, which shows the covariant coupling of fermion of 
spin-$1/2$ to gravity, can be given as
\begin{equation}
{\cal L}=\sqrt{-g}\left(i \, \bar{\psi} \, \gamma^aD_a\psi- m \, \bar{\psi}\psi\right),\label{lag}
\end{equation}
where the covariant derivative and spin connection are defined as
\begin{eqnarray}
D_a=\left(\partial_a-\frac{i}{4}\omega_{bca}\sigma^{bc}\right) \label{cd},
\end{eqnarray}

\begin{eqnarray}
\omega_{bca}=e_{b\lambda}\left(\partial_a e^\lambda_c+\Gamma^\lambda_{\gamma\mu} e^\gamma_c e^\mu_a\right). \label{om}
\end{eqnarray}

We would work in units of $c = \hbar = k_B  = 1$.
We have assumed a torsion-less space-time and the Lagrangian is invariant under the local Lorentz
transformation of the vierbien, $e^a_\mu$, and the spinor field, $\psi(x)$, as
$e^a_\mu(x) \rightarrow \Lambda^a_b(x) e^b_\mu(x)$ and $\psi(x) \rightarrow exp(i \epsilon_{a b}(x) \sigma^{a b}) \psi(x)$, where $\sigma^{bc}=\frac{i}{2}\left[\gamma^a,\gamma^b\right]$ is the 
generator of tangent space Lorentz transformation, the Latin and Greek alphabets indicate
the flat and curved space coordinate respectively. Also
\begin{eqnarray}
e^\mu_a e^{\nu a}=g^{\mu\nu},\hskip0.5cm e^{a \mu} e^b_\mu=\eta^{ab},\hskip0.5cm \{\gamma^a,\gamma^b\}=2\eta^{ab},
\end{eqnarray}
where $\eta^{ab}$ represents the inertial frame of the Minkowski metric and $g^{\mu\nu}$ is the curved space-time metric, here the Kerr geometry.

Now from (\ref{lag}) and (\ref{cd}), it is clear that the product of three
Dirac matrices appears in the Lagrangian and which is
\begin{eqnarray}
\gamma^a\gamma^b\gamma^c=\eta^{ab}\gamma^c+\eta^{bc}\gamma^a-\eta^{ac}\gamma^b
-i\epsilon^{abcd}\gamma^5\gamma_d.
\end{eqnarray}

Thus the spin connection terms are reduced into the combination of an anti-hermitian, 
$\bar{\psi} A_a\gamma^a \psi$, and a hermitian, $\bar{\psi} B^d \gamma^5 \gamma_d \psi$, interaction terms. 
The anti-hermitian interaction term disappears when its conjugate part of Lagrangian is added to 
(\ref{lag}). The only interaction survives in $\cal L$ is the hermitian part and (\ref{lag}) reduces to
\begin{eqnarray}
{\cal L}=det(e)\bar{\psi}\left[(i\gamma^a\partial_a-m)+\gamma^a\gamma^5 B_a\right]\psi,
\label{lagf}
\end{eqnarray}

where
\begin{eqnarray}
B^d=\epsilon^{abcd} e_{b\lambda}\left(\partial_a e^\lambda_c+\Gamma^\lambda_{\alpha\mu} 
e^\alpha_c e^\mu_a\right)\label{bd}
\label{bd}
\end{eqnarray}
and in terms of tetrads, Christoffel connection is reduced as
\begin{eqnarray}
\Gamma^\alpha_{\mu\nu}=\frac{1}{2}\eta^{ij}e^\alpha_i e^\beta_j \left[(e^j_{\beta,\nu}e^p_\mu
+e^j_\beta e^p_{\mu,\nu})\eta_{jp} + (e^j_{\beta,\mu}e^q_\nu+e^j_\beta e^q_{\nu,\mu})\eta_{jq}
- (e^p_{\mu,\beta}e^q_\nu+e^p_\mu e^q_{\nu,\beta})\eta_{pq}\right].
\label{cris}
\end{eqnarray}  
Thus from (\ref{lagf}), the free part of the Lagrangian is, 
${\cal L}_f=det(e)\bar{\psi}\left(i\gamma^a\partial_a-m\right)\psi$, which is exactly same
as the Dirac Lagrangian in the flat space, and the interaction part due to the curvature of
space-time is, ${\cal L}_I=det(e)\bar{\psi}\gamma^a\gamma^5\psi B_a$. It is 
known that Lagrangian for any fermionic field is invariant only under local Lorentz transformation
\cite{bd}. However, if the gravitational four vector field $B_a$, is chosen as constant background in the local frame,
then ${\cal L}_I$ violates CPT as well as particle Lorentz symmetry in the local 
frame.
For example, if $B_a$ is constant and space-like (what we will show later according to Kerr geometry), 
then the corresponding fermion will have different
interactions if its direction of motion or spin orientation changes, 
and thus the breaking of 
Lorentz symmetry in the local frame is natural. 
It should be noted that similar
interaction terms are considered in CPT violating theories and string theory (e.g. \cite{ck}, \cite{kp}).
Here the terms come into the picture automatically, due to the interaction with background curvature,
and therefore the physical origin is very clear. Following \cite{bd}, \cite{ck}, we call the interaction,
${\cal L}_I$, is observer Lorentz invariant but there the particle Lorentz symmetry is broken.
Here, both the kinds of Lorentz symmetry are different obviously as neutrinos are considered
moving under gravitational field and thus they are no longer free. Now ${\cal L}_I$
is CPT violating if it changes sign under CPT transformation. Actually under CPT transformation,
$\bar{\psi}\gamma^a\gamma^5\psi$, which is an axial-vector (pseudo-vector), changes sign.
If $B_a$ does not change sign under CPT, then ${\cal L}_I$
is CPT violating (CPT odd) interaction as well otherwise the interaction is CPT even. 
It is the nature of background metric which determines
whether the functional form of $B_a(x,y,z,t)$ 
is odd (changes sign) under CPT or not. Overall we can say, ${\cal L}_I$ is CPT as well as particle
Lorentz violating interaction (it can be noted that CPT violation necessarily implies the Lorentz
violation in local field theory \cite{greenberg}). 
However, if $B_a$ does not break the symmetry of particle Lorentz transformations in the local
frame, the CPT also cannot be broken.
As we would see, for the
propagation of neutrinos in the Kerr black hole space-times the interaction term
is CPT violating. Thus,  the vector $B_a$ causes breakdown
of Lorentz invariance and  CPT violation.

We would here like to mention that our analysis is different from earlier
studies of interactions violating Lorentz invariance but which were mainly
CPT even \cite{colglash}. These studies were based on interactions
in the flat space-time and thus excluded interactions
of fermions with background gravitational field. The purpose of these
studies was to have  high energy high precision tests of special relativity.
One can then obtain bound on terms in Lagrangian violating Lorentz 
invariance, through various experiments like cosmic ray observations (e.g. \cite{colglash,gio1}), neutrino
oscillations (e.g. \cite{ck,gio2}) etc. We in this paper, focus on the general relativistic
effects on propagation of fermions and we establish that the background
gravitational field plays an interesting role in disguise of vector $B_a$
to cause CPT violation and hence the neutrino$-$anti-neutrino asymmetry. Further,
our analysis, unlike that of \cite{colglash} is based 
on considering interaction terms which violate CPT. As applied to
phenomenology our motivation would be to seek possible generation of
neutrino$-$anti-neutrino asymmetry in the Universe by putting bounds on
parameters of the background black hole space-times. It would be interesting
to extend this analysis to study the phenomenological applications 
e.g. neutrino oscillation as studied earlier \cite{colglash}.

The important factor for our application is that the 
interaction term (${\cal L}_I$) in (\ref{lagf}) 
has different signs for left and right chiral
fields. The coupling term for particles $\psi$ and anti-particles 
$\psi^c$ may be expressed as 
\begin{eqnarray}
\bar{\psi}\gamma^a \gamma^5 \psi= \bar{\psi}_L \gamma^a \psi_L -\bar{\psi}_R \gamma^a \psi_R, 
\label{part}
\end{eqnarray}

\begin{eqnarray}
\bar{\psi}^c \gamma^a \gamma^5 \psi^c = (\bar{\psi}^c)_L \gamma^a (\psi^c)_L 
-(\bar{\psi}^c)_R \gamma^a (\psi^c)_R.  
\label{antipart} 
\end{eqnarray}

Now, if we consider the spinor field as neutrino and since according to the standard model, 
particles (neutrinos) have left chirality and anti-particles (anti-neutrinos) have only right 
chirality, the second term in (\ref{part}) and the first term in (\ref{antipart}) 
will not be present. Thus the spin-connection interaction will have opposite sign for the 
(left-handed) neutrino and the (right-handed) anti-neutrino. Therefore the dispersion
relation of the left and right chirality fields including the
Lorentz and CPT violating term can be written as
\begin{eqnarray}
(p_a \pm B_a)^2=m^2,
\label{dis}
\end{eqnarray}
where the `$+$' and `$-$' signs correspond to the left handed and right handed partners. 

The effect of background gravitational field on the propagation of neutrinos is to
modify the dispersion relation. The vector $B_a$ violates CPT, breaks Lorentz invariance 
and causes the above modification. 

Thus, expanding out (\ref{dis}) for particles ($E_\nu$) and anti-particles ($E_{\bar{\nu}}$) the
dispersion energy becomes
\begin{eqnarray}
\nonumber
E_{\nu} &=& \sqrt{({\vec p} - \vec{B})^2+m^2}+B_0\\
E_{\bar{\nu}} &=& \sqrt{({\vec p} + \vec{B})^2+m^2}-B_0,
\label{edis}
\end{eqnarray}
where we only consider the positive energy solutions. Clearly the energy splitting 
between neutrino and anti-neutrino disappears when the gravitational coupling,
$B_a\longrightarrow 0$ (in the flat space).

An important point to be noted here that while propagating
under a strong gravity neutrinos and anti-neutrinos acquire an effective mass due to the coupling with
the space-time (they no longer are free now). It is this effective mass (which is helicity dependent)
appears in an opposite sign in the Lagrangian with different helicity and
then dispersion relations [see above equation (\ref{edis})].
Neutrinos and anti-neutrinos propagating in gravitational fields would thus have different
energies. This energy difference between particles and anti-particles is the direct result 
of the presence of $B_a$ which violates CPT. We can further evaluate the difference in number
density of neutrinos and anti-neutrinos propagating in a gravitational background as
\begin{eqnarray}
\Delta n=\frac{g}{(2\pi)^3}\int_{R_i}^{R_f} dV \int d^3 |{\vec p}|
\left[\frac{1}{1+exp(E_{\nu}/T)}-\frac{1}{1+exp(E_{{\bar{\nu}}}/T)}\right],
\label{fn}
\end{eqnarray}
where $R_i$ and $R_f$ refer to two extreme points of the interval over which the
asymmetry is measured
and $dV$ is the small volume element in that interval.

Here the above modifications to dispersion relations and then the asymmetry 
is something similar to
the effect on baryogenesis of spontaneous CPT violation in string based
scenario \cite{b97}.
In that case fermions are considered with extra interactions
responsible for different chemical potentials and the asymmetry between quarks
and anti-quarks. Here a similar interaction originates inherently to
bring the neutrino-antineutrino asymmetry. 

In the case, when $B_0$ is vanishing, the integrand is an odd function
and hence $\Delta n = 0$. Any non zero value of $B_0$ would yield a $\Delta n 
\neq 0$ and hence neutrino asymmetry. Thus to create any neutrino asymmetry, a non-zero
$B_0$ is required, and it does not matter whether  $B_i$s ($i=1,2,3$) are vanishing 
or not. This is the reason, why the metric 
should have a non-zero off-diagonal spatial components for the neutrino asymmetry to occur.

\section{Neutrino asymmetry around black holes} 

An example of origin of the neutrino asymmetry in a black hole space-time 
can be given for the Kerr geometry. For simplicity of analysis we would 
write the Kerr metric in Cartesian-like
form, i.e., our variables are $t (=x_0),x (=x_1),y (=x_2),z (=x_3)$. 
We would however here stress that the conclusions are independent of the choice of
coordinate system to describe the background space-time, as we comment
in \S 4. In the Cartesian form, the Kerr metric 
with signature $[+ - - -]$ can be written as \cite{doran}
\begin{equation}
d s^2 = \eta_{ij} \, d x^i \, d x^j - \bigg[ \frac{2 \alpha}{\rho} \, s_i \, v_j + \alpha^2 \, v_i \, v_j \bigg] d x^i \, d x^j \label{met}
\end{equation}
where 
\begin{equation}
\alpha = \frac{\sqrt{2 M r}}{\rho}, ~~~~ ~~~~ \rho^2 = r^2 + \frac{a^2 z^2}{r^2},
\end{equation}
\begin{equation}
s_i = \left(0, ~~ \frac{r x}{\sqrt{r^2 + a^2}}, ~~ \frac{r y}{\sqrt{r^2 + a^2}}, ~~ \frac{z \sqrt{r^2 + a^2}}{r} \right),
\end{equation}
\begin{equation}
v_i = \left(1, ~~ \frac{a y}{a^2 + r^2}, ~~ \frac{- a x}{a^2 + r^2}, ~~ 0 \right) .
\end{equation}

Here $a$ and $M$ are respectively the specific angular momentum and mass of the Kerr
black hole and $r$ is positive definite satisfying the following equation,
\begin{equation}
r^4 - r^2 \, \left(x^2 + y^2 + z^2 - a^2 \right) - a^2 z^2 = 0.
\end{equation}

The corresponding non-vanishing component of tetrads (vierbiens) are \cite{doran} 
\begin{eqnarray}
\nonumber
&&e^0_t = 1, ~~ ~~  e^1_t = - \frac{\alpha}{\rho} \, s_1, ~~ ~~ e^2_t = - \frac{\alpha}{\rho} \, s_2, ~~ ~~ e^3_t = - \frac{\alpha}{\rho} \, s_3, \label{tet1}  \\
\nonumber
&&e^1_x = 1 - \frac{\alpha}{\rho} \, s_1 \, v_1, ~~ ~~ e^2_x = - \frac{\alpha}{\rho} s_2 \, v_1, ~~ ~~ e^3_x = - \frac{\alpha}{\rho} \, s_3 \, v_1, \label{tet2} \\
&&e^1_y = - \frac{\alpha}{\rho} \, s_1 v_2, ~~ ~~ e^2_y = 1 - \frac{\alpha}{\rho} \, s_2 \, v_2, ~~ ~~ e^3_y = - \frac{\alpha}{\rho} \, s_3 \, v_2, ~~ ~~ e^3_z = 1 - \frac{\alpha}{\rho} \, s_3 \, v_3 \label{tet3}.
\label{tetnonv}
\end{eqnarray}

Using (\ref{bd}), (\ref{cris}), (\ref{met}) and (\ref{tetnonv}), the 
gravitational scalar potential can be evaluated as 
\begin{eqnarray}
 B^0 &=& e_{1\lambda}  \left( \partial_3 e_2^\lambda - \partial_2 e_3^\lambda
 \right) + e_{2 \lambda} \left( \partial_1 e_3^\lambda - \partial_3 e_1^\lambda
 \right) + e_{3 \lambda} \left( \partial_2 e_1^\lambda - \partial_1 e_2^\lambda
 \right).
\label{bo}
\end{eqnarray} 
Similarly, the gravitational vector potentials $B^1,B^2,B^3$ can be calculated.  
From (\ref{bo}), it is clear that $B_0$ will become zero, if all the 
off-diagonal spatial components of the metric are zero (i.e. $g_{ij}=0$,
where, $i\neq j\rightarrow 1,2,3$). In other words we can say, 
there should be a minimum space-space curvature coupling effect to give rise to a nonzero scalar 
potential, $B^0$.

One can easily check from (\ref{bo}) along with (\ref{tetnonv}) that under CPT transformation,
form of $B_0$ would not behave as odd function,
more precisely, $B_0$ neither flips its sign 
[$B_0(-x,-y,-z,-a,M) \neq -B_0(x,y,z,a,M)$] nor be invariant [$B_0(-x,-y,-z,-a,M) \neq B_0(x,y,z,a,M)$]. 
The same would hold for $B_1,B_2,B_3$. Therefore, $B_a$ leads to CPT violation
in the Lagrangian.
As mentioned earlier, this nature of $B_a$ under CPT totally depends on the choice of
background metric, the space-time, where the neutrino is propagating. A case of the space-time 
was studied earlier \cite{mmp02} where $B_0$ 
flips its sign (odd function) under CPT and thus overall ${\cal L}_I$ is CPT invariant. 
However, the present case where the space-time is chosen around a rotating black hole
gives rise to an actual CPT and Lorentz violating situation.

Now we will show, how does the above mentioned 
property of neutrino, along with the effect of curvature,
generates its asymmetry. For simplicity, let us consider a special case of a black
hole space-time with $\vec B \, . \, \vec p \, \ll \, B_0 \, p^0$ and the black hole curvature
effect is such that $B_a \, B^a$ term can be neglected, and thus only the
first order curvature effect is important. Then in the ultra-relativistic 
regime, we get from (\ref{fn}),
\begin{eqnarray}
\Delta n=\frac{g}{(2\pi)^2} \, T^3 \, \int_{R_i}^{R_f} \, \int_0^\infty \, \int_0^\pi \, 
\left[\frac{1}{1+ e^u \, e^{ B_0/T}}-\frac{1}{1+ e^u \, e^{-  B_0/T}}\right]
\, u^2 \, d\theta \, du \, dV
\label{fn1}
\end{eqnarray} 
where $u = |{\vec p}|/T$. If $B_0 \ll T$, then 
\begin{equation}
\Delta n \sim g \, T^3 \, \left(\frac{\overline{B_0}}{T} \right),
\label{nas}
\end{equation}
$\overline{B_0}$ indicates the integrated value of $B_0$ over the space.

It should be noted that the sign of above asymmetry would depend on the
overall sign of $\overline{B_0}$, which depends on 
details of mass and angular momentum of the black hole.
A large asymmetry can be achieved in practical situations as in accretion disks
and case of Hawking radiation bath. In the first case, the virial temperature of
thermal bath for the neutrinos can be as high as $10^{12}$ K $\sim 100$ MeV.
Therefore, to have a neutrino asymmetry around a Kerr
black hole, the space-time curvature effect has to be at least one order of
magnitude weaker, say, $\overline{B_0}\leq 10$ MeV, than the energy of bath.
Moreover, the phenomena of a Hawking bath looks very promising, where
small primordial black holes {\bf are} produced in copious amounts. We know, all the 
primordial black holes of mass less than $10^{15}$ gm have been evaporated 
already. Only black holes of mass, $M > 10^{15}$ gm,  still exist today.
The temperature of Hawking bath can be given as
\begin{equation}
T=\frac{\hbar}{8\pi k_B M}\sim 10^{-7} K \left(\frac{M_\odot}{M}\right).
\label{ht}
\end{equation}
Thus the primordial black hole of masses of the order $10^{15}$ gm can generate Hawking temperature of 
the order $T\sim 10^{11}$ K $\sim 10$ MeV. Hence, to generate a neutrino asymmetry, the restriction on 
curvature effect should be, $\overline{B_0}\le 1$ MeV. If we consider, temperature of bath, 
$T\sim 10^{11}$ K $\sim 1.6\times 10^{-5}$ erg, $\overline{B_0}\sim 1.6\times 10^{-6}$ erg, then
$\Delta n\sim 10^{-16}$. If there are typically $10^6$ number of black holes with same sign of
$\overline{B_0}$, $\Delta n\sim 10^{-10}$,
which agrees with the observed neutrino asymmetry in the Universe.

\section{Conclusion}

We have proposed a new mechanism to generate the neutrino asymmetry in the present epoch of 
the Universe. Such a mechanism can provide the neutrino 
asymmetry in addition to the relic neutrino asymmetry arising due to the
leptogenesis in the early Universe. We have explicitly demonstrated this through
an example where neutrinos are propagating around Kerr black holes. 
Here, for convenience, we have chosen the Kerr metric in Cartesian-like coordinates $(x,y,z,t)$.
It is seen that, in presence of any off-diagonal spatial component of the metric
($g_{ij}, i\ne j\rightarrow 1,2,3$) the scalar potential part ($B_0$) of the 
space-time interaction is non-zero. According to the present mechanism, this
scalar potential is actually responsible for the neutrino asymmetry in the Universe.
If all the $g_{ij}$s are zero, $B_0$ vanishes and hence $\Delta n=0$.
Although, this restriction on $g_{ij}$ as well as $B_0$, to have a non-zero neutrino asymmetry, 
is made here on the basis of a fixed coordinate system,
in principle we can choose any other kind of coordinate system
to describe the background geometry and to generate neutrino asymmetry. 
One can easily check that, in the Boyer-Lindquist
coordinate system \cite{bl}, $B_0$ is zero. But in that case, at least one 
non-zero $B_i$ ($i\rightarrow 1,2,3 \equiv\rho,\theta,\phi$) is required
 i.e., for example, presence of $g_{03}$ is enough, to
give rise to neutrino asymmetry. Thus the restriction to generate the neutrino asymmetry
around the black hole is that the black hole must be rotating and hence the system
is symmetric axially.

The asymmetry can be produced in accretion disks or/and
Hawking radiation baths, which provide high enough temperature for such an effect to occur. 
Assume that, there are $N_i$ number of $i$-type black holes in Universe, each producing a net 
curvature effect ${B_0}_i$ in a typical temperature of the system $T_i$, then the neutrino asymmetry due to the 
presence of a black hole of kind-$i$ can be given as
\begin{equation}
\Delta n_i=10^{-10}\left(\frac{N_i}{10^6}\right)\left(\frac{{B_0}_i}{10^{-6} erg}\right)
\left(\frac{T_i}{10^{-5} erg}\right)^2.
\label{neuasy}
\end{equation}
If all the black holes in Universe are of $i$-kind and there are $10^6$ such black holes, 
the curvature effect and temperature of the system
are $10^{-6}$ erg and $10^{-5}$ erg respectively, then the neutrino asymmetry in Universe 
is $10^{-10}$. Any change of $N_i$, ${B_0}_i$ and $T_i$ will affect $\Delta n$. In
general the net neutrino asymmetry in the Universe can be written as
\begin{equation}
\Delta n=\sum_i \Delta n_i,
\label{netuasy}
\end{equation}
where ${B_0}_i$ and then $\Delta n_i$ can be positive as well as negative depending on the kind of 
black holes (parameters of background space-times). However, it is difficult
to predict about total number of black holes along with the variation of all 
their parameters. Therefore to estimate the total neutrino asymmetry
due to black holes may not be possible physically at this stage. Moreover
depending on the sign of ${B_0}_i$s, we can not say whether the net effects of
asymmetry will cancel out completely or not.
 
It should be reminded that, this kind of the neutrino asymmetry can be achieved
in some other space-time geometry where $B_0$ is non-vanishing. 
The Kerr geometry is chosen as an example only in the 
present paper. However, as the number of black hole may be very high and the physics
behind it is very well established, it is advantageous to consider  
black hole space-times to built up a
real feeling about the physics behind this new mechanism. Also the advantage to deal 
with Cartesian-like coordinate system is that the structure of the Dirac gamma matrices 
($\gamma^0$, $\gamma^i$) are very well known there. 

In our earth, the curvature effect is measured as $10^{-34}$ MeV$\sim$ $10^{-40}$ erg \cite{mmp02} and
the temperature is about $10^{-14}$ erg $\sim 10^{-2}$ eV. Thus, according to 
(\ref{neuasy}), the neutrino asymmetry comes out as $10^{-68}$ which is too small effect
to observe. However, in earth's laboratory, neutrinos can be examined in a high temperature 
bath. As the asymmetry is proportional to the square of temperature,
it can be enhanced by increasing temperature in the laboratory. 
Moreover, if there are large
number of earth like systems exist in the Universe, overall $\Delta n$ may also increase
according to (\ref{netuasy}).

Similar modifications
to the dispersion relations may arise for virtual black holes also. The
emergence of a birefringence effects associated with quantum gravity
corrections, have already been seen. The modification of dispersion relation
due to the quantum gravitational medium effect was shown, first for the photon
as a helicity independent manner \cite{gio3} and then in a helicity dependent
way \cite{gp}. Those modifications also appear as CPT and Lorentz violation.
Those works are involved with the space-time curvature and Maxwell's field mainly.
If one carefully compares those with our present one, it comes that the fundamental
origin of the effects are same! In our case, it is the space-time curvature
coefficients, which couples with fermions to produce the effect, i.e. the
modification to the dispersion relations. In addition,
here we bring the neutrino asymmetry based on this gravity effect.
That asymmetry only will survive if the space-time deviates from spherical
symmetry. Otherwise, say for the expanding FRW cosmology \cite{gio3}, though
the dispersion relations get modified but the asymmetry does
appear (as explained the reasons in various places in the above text).
Point is, whatever be the origin of space-time curvature (due to presence of
black holes or expanding era of early Universe with primordial fluctuations etc.)
propagating neutrinos in curved space-time always produce this effect.

Our mechanism essentially works in the 
 presence of a pseudo-vector term ($\bar{\psi}\gamma^a\gamma^5\psi$) multiplied by a background
curvature coupling ($B_a$).
This is the CPT and Lorentz violating term, which picks up an opposite sign between a neutrino
and an anti-neutrino. 
Thus the CPT violating nature of the gravitational
interaction with spinor is an essential condition in success of the mechanism.
Thus we propose, to generate the neutrino asymmetry in presence of gravity, following
criteria have to be satisfied as: (i) The space-time should be axially symmetric,
(ii) the interaction Dirac Lagrangian must have a CPT violating term which may be an
axial-four vector (or pseudo-four vector) multiplied by a curvature coupling four vector potential.
(iii) the temperature scale of the system should be large with respect
to the energy scale of the space-time curvature.
If all these conditions are satisfied simultaneously, our mechanism will give
rise to the neutrino asymmetry in Universe. It would be interesting to explore the
further theoretical and phenomenological consequences of the role of 
background gravitational curvature for neutrinos, which might offer new
insights in the interplay of gravity and standard model interactions and
specially of neutrino physics.

\vskip1cm
\noindent{\large Acknowledgment}\\

Author thanks D. V. Ahluwalia-Khalilova for discussions from two years ago in IUCAA, India. 
Thanks are also directed to V. A. Kosteleck\'y for his comments and suggestions.
This work was supported in part by NSF grant AST 0307433.


\begin{references}

\bibitem{admcdonald} I. Affleck, M. Dine, Nucl. Phys. {\bf B 249}, 361 (1985);
J. McDonald, Phys. Rev. Lett {\bf 16}, 748 (2000). 

\bibitem{cosmoeffects}  J. Lesgourgues, S. Pastor, Phys. Rev. {\bf D60}, 
103521, (1999); S. Hannestad, Phys. Rev. Lett. {\bf 85}, 4203 (2000); 
J. P. Kneller et al., Phys. Rev. {\bf D64}, 123506 (2001); S. H. Hansen et al.,
Phys. Rev. {\bf D65}, 023511 (2002); A. D. Dolgov, Phys. Rep. {\bf 370}, 333 (2002); M. Orito et al., Phys.Rev. {\bf D65} 
123504, (2002).

\bibitem{fargion} D. Fargion, M. Grossi, P. Lucentini, \& C. Di Troia, Supplement B, J. Phys. Soc. Jpn. 
{\bf 70}, 46 (2001); D. Fargion \& M. Khlopov, Astropart. Phys. {\bf 19}, 441 (2003); 
D. Fargion, in {\it Particle Astrophysics Instrumentation}, Ed. W. Gorham, Proceedings of the SPIE
vol.4858, p.1 (2003); hep-ph/0208093.

\bibitem{weiler} T. Weiler, Astropart. Phys. {\bf 11}, 303 (1999);
A. Kusenko \& T. Weiler, Phys. Rev. Lett. {\bf 88}, 161101 (2002).

\bibitem{gzk} K. Greisen, Phys. Rev. Lett. {\bf 16}, 748 (1966);
G. T. Zatsepin, V. A. Kuzmin, Pis\'ma Zh. Eksp. Teor. Fiz. {\bf 4}, 114 (1966)
[JETP Lett. {\bf 4}, 78 (1966)];
G. Gelmini, A. Kusenko, Phys. Rev. Lett. {\bf 82}, 5202 (1999).  

\bibitem {pap51} A. Papapetrou, Proc. Roy. Soc. Lond. {\bf A209}, 248 (1951).

\bibitem {dixon} W. G. Dixon, Phil. Tran. R. Soc. Lond {\bf A277}, 59 (1974).

\bibitem{ads} J. Anandan, N. Dadhich, P. Singh, Phys. Rev. {\bf D68},
124014 (2003); Int. J. Mod. Phys. {\bf D12}, 1651 (2003).

\bibitem {jeeva} J. Anandan, Phys. Lett. {\bf A 195}, 284 (1994).

\bibitem {m00} B. Mukhopadhyay, Class. Quantum Grav. {\bf 17}, 2017 (2000).

\bibitem {p71} L. Parker, Phys. Rev. {\bf D3}, 346 (1971).

\bibitem {chandra76} S. Chandrasekhar, Proc. Roy. Soc. Lond. {\bf A349}, 571 (1976).

\bibitem {p80} L. Parker, Phys. Rev. {\bf D22}, 1922 (1980).

\bibitem {lalak} Z. Lalak, S. Pokorski, \& J. Wess, Phys. Lett. {\bf B355}, 453 (1995).


\bibitem {bd} N. Birrel, \& P. Davis in {\it Quantum fields in curved space}, 
Cambridge: Cambridge University Press (1982).

\bibitem {sw} J. Schwinger in {\it Particles, sources and fields}, 
Redwood City, California: Addison-Wesley Publishing Company, INC. (1973).

\bibitem {mmp02} S. Mohanty, B. Mukhopadhyay, \& A. R. Prasanna, Phys. Rev {\bf D65 }, 122001 (2002).

\bibitem {ck} D. Colladay, \& A. Kosteleck\'y, Phys. Rev. {\bf D55}, 6760 (1997); Phys. Rev. {\bf D58}, 116002 (1998).

\bibitem {d99} H. Dehmelt et al., Phys. Rev. Lett. {\bf 83}, 4694 (1999).

\bibitem {bklr} R. Bluhm, A. Kosteleck\'y, C. Lane, \& N. Russell,
Phys. Rev. Lett. {\bf 88}, 090801 (2002).

\bibitem {mpla} P. Singh, \& B. Mukhopadhyay, Mod. Phys. Lett. {\bf A 18}, 779 (2003).

\bibitem {pram} B. Mukhopadhyay, \& P. Singh, Pramana {\bf 62}, 775 (2004).

\bibitem {ahl1} D. V. Ahluwalia-Khalilova, Int. J. Mod. Phys. {\bf D 13}, 2361 (2004).

\bibitem {ahlbur} D. V. Ahluwalia, \& C. Burgard, Gen. Rel. Grav. {\bf 28}, 1161 (1996).

\bibitem {ahl2} D. V. Ahluwalia, Gen. Rel. Grav. {\bf 29}, 1491 (1997). 

\bibitem {kk} K. Konno, \& M. Kasai, Prog. Theor. Phys. {\bf 100}, 1145 (1998).

\bibitem {wud} J. Wudka, Phys. Rev. {\bf D64}, 065009 (2001).

\bibitem {kosgrav} V. A. Kosteleck\'y, Phys. Rev. {\bf D 69}, 105009 (2004).

\bibitem {kp} V. A. Kosteleck\'y, \& R. Potting, Phys. Rev. {\bf D51}, 3923 (1995).

\bibitem {greenberg} O. W. Greenberg, Phys. Rev. Lett. {\bf 89}, 231602 (2002).

\bibitem {colglash} S. Coleman, \& S. L. Glashow, Phys. Rev. {\bf D59}, 116008 (1999).

\bibitem {gio1} G. Amelino-Camelia, \& T. Piran, Phys. Rev. {\bf D64}, 036005 (2001).

\bibitem {gio2} G. Amelino-Camelia, Int. J. Mod. Phys. {\bf D12}, 1633 (2003).

\bibitem {b97} O. Bertolami, D. Colladay, V. A. Kostelecky \& R. Potting, Phys. Lett. {\bf B395}, 178 (1997).






\bibitem{doran} C. Doran, Phys. Rev. {\bf D61}, 067503 (2000).

\bibitem{bl} R. H. Boyer, \& R. W. Lindquist, J. Math. Phys. {\bf 8}, 265 (1967).

\bibitem{gio3} G. Amelino-Camelia, J. Ellis, N. E. Mavromatos, D. V. Nanopoulos, \& S. Sarkar,
Nature {\bf 393}, 763 (1998).

\bibitem{gp} R. Gambini, \& J. Pullin, Phys. Rev. {\bf D59}, 124021 (1999).


\end{references}
\end{document}